\documentclass[10pt]{iopart}

\usepackage{graphicx}
\usepackage{gensymb}
\usepackage{Citesort}
\usepackage{color}

\begin{document}

\title{Rydberg States of Helium in Electric and Magnetic Fields of Arbitrary Relative Orientation}

\author{Ond\v rej Tk\'a\v c, Matija \v Ze\v sko, Josef A. Agner, Hansj\"urg Schmutz, Fr\'ed\'eric Merkt }

\address{Laboratory of Physical Chemistry, ETH Zurich, CH-8093 Zurich, Switzerland}
\ead{frederic.merkt@phys.chem.ethz.ch}
\vspace{10pt}
\begin{indented}
\item[]December 2015
\end{indented}

\begin{abstract}
A spectroscopic study of Rydberg states of helium (\it n \rm = 30 and 45) in magnetic, electric and combined magnetic and electric fields with arbitrary relative
orientations of the field vectors is presented. The emphasis is on two special cases where (i) the diamagnetic term is negligible 
and both paramagnetic Zeeman and Stark effects are linear ($n$ = 30, $B \leq$  120 mT and $F$ = 0 - 78 V/cm ), and (ii) the diamagnetic term is dominant and the Stark effect is linear 
($n$ = 45, $B$ = 277 mT and $F$ = 0 - 8 V/cm). Both cases correspond to regimes where the interactions induced by the electric and magnetic fields are much weaker than the Coulomb interaction, 
but much  stronger than the spin-orbit interaction. The experimental spectra are compared to spectra calculated by determining the eigenvalues of the Hamiltonian matrix describing helium Rydberg states
 in the external fields. The  spectra and the calculated energy-level diagrams in external fields reveal avoided crossings between levels of different $m_l$ values and pronounced $m_l$-mixing effects  
at all angles between the electric and magnetic field vectors other than 0. These observations are discussed in the context of the development of a method to generate dense samples 
of cold atoms and molecules in a magnetic trap following Rydberg-Stark deceleration.
\end{abstract}

%
\vspace{2pc}

\submitto{Journal of Physics B-Atomic Molecular and Optical Physics}
%
\maketitle 
\ioptwocol

\section{Introduction}
The investigation of the effects of external magnetic and electric fields on atomic spectra is an important activity in atomic physics. Many 
theoretical and experimental studies have been made to understand the energy-level structure of Rydberg atoms in magnetic, electric and combined electric and magnetic 
fields \cite{gallagher94a,stebbings83a}. However, most studies were performed for only one of the two fields or for the special cases of parallel or perpendicular 
orientations of the electric and magnetic field vectors. 
Many studies of the effects of magnetic fields were performed using strong fields, for which the diamagnetic contribution is dominant, starting with the early work 
of Jenkins and Segr\'e \cite{jenkins39a} and the discovery of quasi-Landau resonances by Garton and Tomkins \cite{garton69a}. Zimmerman \it et al. \rm
\cite {zimmerman78a} studied the diamagnetism of Na Rydberg states by high-resolution spectroscopy and analyzed in detail the magnetic-field 
dependence of the level structure from the low to the high field regime. The effects of high magnetic 
fields were also explored experimentally in the Rydberg states of other atoms, including barium \cite{droungas95a}, helium \cite{veldt92a}, rubidium \cite{economou78a} and 
hydrogen in the quasi-Landau regime \cite{holle86a}. 

The pure Stark effect was also studied in numerous Rydberg atoms, including the alkali and alkaline-earth metal 
atoms \cite{zimmerman78b,zimmerman79a,braun86a,grebenkin86a} and the rare gases \cite{lahaye89a,ernst88a,fielding92a,gruetter08a} and in molecules \cite{bordas87a,chevaleyre86a,fielding91a,fielding94a}. 

Early experiments on Rydberg states in parallel electric and magnetic fields were performed in lithium \cite{cacciani88a,cacciani88b,cacciani88c}, hydrogen \cite{cacciani92a} and 
helium \cite{veldt93a}. These studies also presented complete analyses of the observed structures, from which resulted a detailed understanding of the relevant physical processes and 
a classification of the Rydberg states according to different types of behaviour in the fields. In all cases, excellent agreement between measured and calculated spectra was obtained. Recently, 
the Rydberg spectrum of Rb was studied in the presence of strong magnetic and weak parallel electric fields in the $n$-mixing regime, 
with the goal of preparing states of large dipole moments and large optical excitation cross sections for possible applications in quantum information processing \cite{paradis13a}. 

An experimental study of Rydberg states of H in strong perpendicular magnetic and electric fields was reported by Weibusch \it et al. \rm \cite{weibusch89a}. 
The effect of perpendicular electric and magnetic fields was also examined in Rydberg states of rubidium \cite{penent84a,penent88a}, sodium \cite{korevaar83a} and barium \cite{abdulla04a,connerade05a}. 
Rydberg atoms in strong perpendicular electric and magnetic fields are of interest because they exhibit a potential energy surface for the electron motion that has two minima \cite{burkova76a,bhattacharya82a}. 
A narrow and deep potential well is centered at the nucleus and arises from the Coulomb interaction, whereas a shallower well occurs at large distances from the nucleus. 
The first experimental evidence of a field-induced potential minimum at large distances was reported by Fauth \it et al. \rm \cite{fauth87a} following observation of the large electric dipole moment 
associated with these states. Rydberg atoms in strong perpendicular electric and magnetic fields are also of interest for studying the quantum-mechanical properties of systems for which 
the corresponding classical behaviour is chaotic \cite{milczewski94a,milczewski97a}, which is the case when the interactions with the external fields become comparable in strength to the Coulomb 
interaction \cite{raithel91a,raithel93a}. 
 
Only a few theoretical studies have been devoted to the behaviour of Rydberg atoms in electric and magnetic fields with arbitrary relative orientations. 
Calculations for weak external fields using perturbation 
theory were reported in \cite{pauli26a,solovev83a} and the effects of strong electric and magnetic fields were studied in \cite{fassbinder96a,melezhik93a}. In their theoretical study 
of the hydrogen atom in combined electric and magnetic fields with arbitrary relative orientations, Main \it et al. \rm\cite{main98a} predicted large avoided crossings 
between eigenstates of the same approximately conserved $n$ values, a phenomenon which was 
interpreted as a quantum manifestation of intramanifold chaos.   

Our motivation to study the spectrum of Rydberg states of He in electric and magnetic fields with arbitrary relative orientations originated in experiments in which we seek to develop a new 
trap-loading scheme for cold paramagnetic atoms and molecules in supersonic beams relying on Rydberg-Stark deceleration and trapping \cite{hogan08a}. 
The  strategy we follow to increase the density of cold trapped atoms and molecules consists of first decelerating and deflecting the Rydberg atoms using a Rydberg-Stark 
decelerator \cite{hogan08a,seiler11a,hogan11a}, loading the atoms in an off-axis electric trap and waiting for them to radiatively decay to the ground or a metastable state, i.e., the 1s2s $^3$S$_1$ = 2 $^3$S$_1$ state 
in the case of He. Superimposing a magnetic trap on the electric trap would enable us to increase the density of trapped atoms at each cycle of our experimental procedure. A similar strategy, 
but based on multistage Stark deceleration rather than Rydberg-Stark deceleration, has been pursued to trap NH molecules \cite{riedel11a}.

Figure 1 shows the distributions of electric (red arrows) and magnetic (blue arrows) fields in the overlaid electric and magnetic traps. The electric trap is generated by four electrodes 
in quadrupole configuration, and the magnetic trap, with a field minimum at the centre of the electric trap, is created by two permanent magnets placed above and 
below the plane of the figure.  
Atoms trapped in such electric- and magnetic-field distributions experience all possible relative orientations between the electric-
and magnetic-field vectors. Whereas the weak electric fields we use for trapping ($F < 250$ V/cm, see figure 1)  will not influence the behaviour 
of the ground state or metastable states accumulating in the trap, one can anticipate that the magnetic fields of up to 50 mT will influence the Rydberg-Stark deceleration and trapping processes. 
The success of Rydberg-Stark deceleration and trapping experiments critically depends on the ability to carry out realistic particle-trajectory simulations and these in turn require a good knowledge of the field dependence of the energy levels. The purpose of the study presented in this article is to develop and validate efficient procedures to compute the energy-level structure of Rydberg atoms (and molecules) in electric and magnetic fields of arbitrary relative orientation. Spectroscopic studies of Rydberg states of He in magnetic, electric and combined 
magnetic and electric fields with arbitrary relative orientations are presented here as a necessary step in our overall strategy.

Procedures to compute Rydberg spectra in pure electric fields, pure magnetic fields, and parallel and perpendicular electric and magnetic fields are well established, as reviewed above. We nevertheless chose to present such spectra because (1) pure Stark and Zeeman spectra are essential to calibrate the field strengths by comparison between experimental and computed spectra, (2) the field strengths themselves are crucial to accurately set the angle between electric and magnetic fields, and (3) spectra recorded for parallel and perpendicular electric and magnetic fields are useful, as limiting cases, in the validation of the procedure devised to calculate spectra for arbitrary angles between the electric- and magnetic-field vectors.

\begin{figure}
 \includegraphics{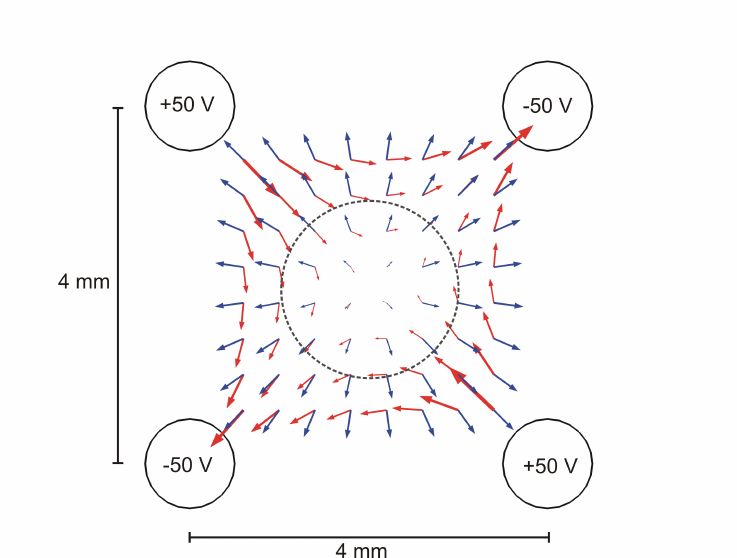}
 \caption{\label{} Schematic diagram of the overlaid electric and magnetic quadrupole trap with the magnetic (blue arrows) and electric (red arrows) field distribution in the central plane. The electric trapping fields are generated by applying $\pm 50$ V to the four parallel cylindrical metallic rods used as electrodes in a quadrupolar arrangement. The quadrupolar magnetic-field distribution is created by two permanent magnets in north-north configuration positioned in the center of the four electrodes above and below the plane of the figure, as indicated by the dashed circle. The maximal sizes of the red and blue arrows correspond to 250 V/cm and 47 mT, respectively.
}
 \end{figure} 

At the $n$ values around 30 used and the typical fields of 0-250 V/cm and 0-50 mT relevant for our planned trapping schemes, the Stark and Zeeman interactions 
are much below the range where quantum chaos is expected. These interactions can be adequately treated by perturbation theory, which is the approach we follow in the 
analysis of our spectra. Because adiabatic transitions between Rydberg states of different electric dipole moments have an adverse effect on the deceleration and trapping efficiency, emphasis 
is placed on the characterization of avoided crossings between the Rydberg states in energy-level maps in which the energy eigenvalues are plotted as a function of the electric field, the magnetic 
field, and the angle between the field vectors.  

This paper is organized as follows: Section 2 contains the descriptions of the experimental setup and the method of calculating the spectra of Rydberg helium in 
external fields. The experimental spectra are presented and discussed in Section 3, where they are also compared with calculated spectra. The discussion starts 
with the simplest case of Rydberg helium spectra measured in pure magnetic and pure electric fields, continues with the special cases of parallel and perpendicular fields and ends with 
the general case of fields with arbitrary relative orientation. A brief conclusion is given in Section 4.  

\section{Methods}
\subsection{Experimental setup}
A schematic view of the experimental setup is presented in figure 2. A cold supersonic beam of helium is formed by expanding pure He gas into vacuum from a reservoir held at a 
stagnation pressure of 2.5 bar using a pulsed valve. The valve is cooled to 130 K with liquid nitrogen, resulting in a beam velocity of 1200 m/s.   
Triplet 2 $^3$S$_1$ helium (called metastable He or He* hereafter) is produced in an electric discharge in the high-pressure region at the exit of the nozzle, 
as described in \cite{allmendinger13a}. After passing a skimmer, the He* atoms enter a photoexcitation and ionization region surrounded by four parallel cylindrically-symmetric 
electrodes and by two pairs of coils in Helmholtz configuration. The separation between the outer two electrodes in the stack is 1.5 cm and the four electrodes are equally spaced. 
The inner two electrodes are used to ensure the homogeneity of the electric field in the photoexcitation region. To generate magnetic fields perpendicular (parallel) to the electric field, two 
coils are used with a center-to-center distance of 50 mm (33 mm) and inner and outer radii of 84 (44.5) and 114.6 (75.1) mm, respectively. The number of windings of each coil is 70.   

\begin{figure*}
 \includegraphics{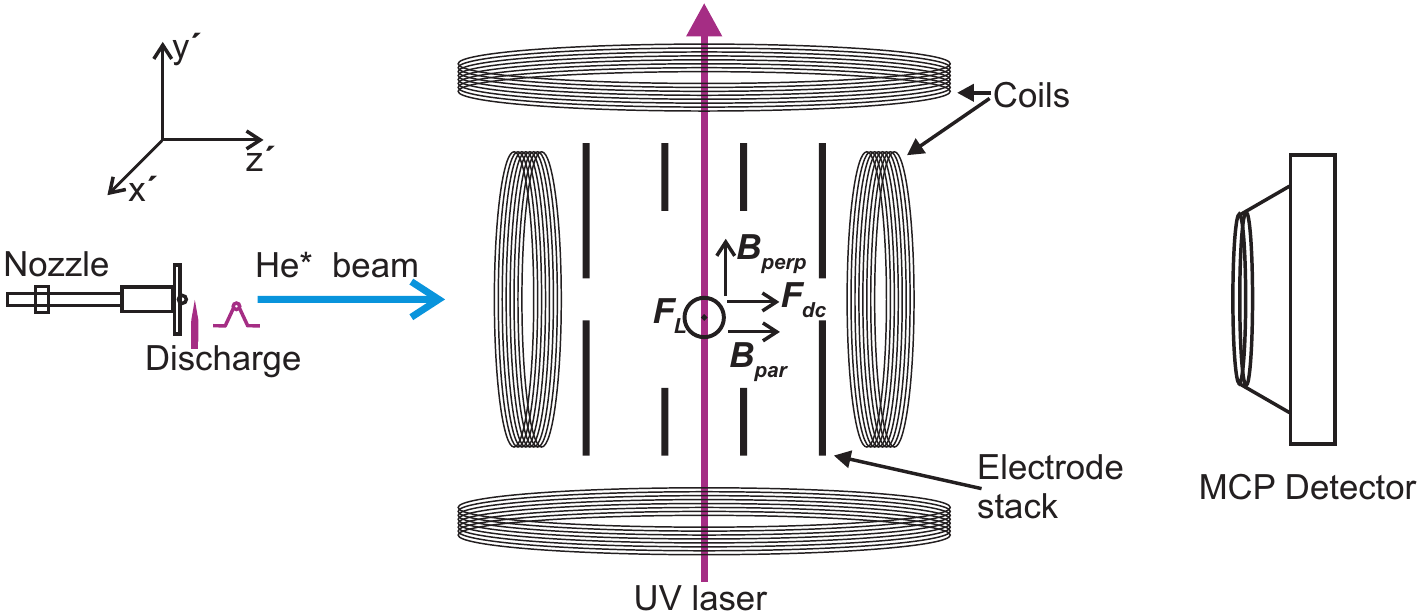}
 \caption{\label{} Schematic diagram of the photoexcitation region displaying the electrode stack used to generate the electric field and the two solenoid pairs used to produce 
the magnetic field. The directions of the atomic beam, the electric field ($\vec{F}_{\rm{dc}}$), the laser beam polarization ($\vec{F}_{\rm{L}}$) and the magnetic fields ($\vec{B}_{\rm{par}}$) 
and ($\vec{B}_{\rm{perp}}$) generated by the two pairs of solenoids are indicated.}
 \end{figure*} 

The He* atoms are excited to Rydberg states using a pulsed (repetition rate 25 Hz, pulse duration $\sim$3 ns, pulse energy 150 $\rm{\mu}$J) narrow-band (full width at half 
maximum 150 MHz) UV laser with a wavelength tunable in the region between 260.41 and 260.88 nm for transitions from the 2 $^3$S$_1$ state to Rydberg states with $n$ between 30 and 45. To generate the UV laser radiation, a 
continuous-wave single-mode tunable IR diode laser (power 30 mW, wavelength 782 nm) is pulse amplified using three successive dye cells operated with the dye Styryl 11 and pumped with the second 
harmonic (532 nm) of a Nd:YAG laser. The output (1 mJ/pulse) of the dye amplification stages is frequency doubled with a BBO crystal, and the doubled 
light, with a wavelength of $\sim$391 nm, is mixed in another BBO crystal with the fundamental IR laser output to get the desired UV radiation. 
The vacuum wavenumber of the UV laser radiation is 
determined from the wavenumber of the IR radiation, which is
measured using a wavemeter (accuracy 3$\sigma \simeq$ 200 MHz). The He Rydberg atoms (\it n \rm = 30 or 45)  
are detected by field ionization with a pulsed electric field of 2 kV/cm, which also extracts the He$^+$ ions toward a detector consisting of a pair of microchannel plates in chevron configuration. 
The joint effects of the laser bandwidth and Doppler broadening led to single transitions having full widths at half maximum of 0.007 cm$^{-1}$. Consequently, all calculated stick spectra were convoluted with a Gaussian line-shape function of 0.007 cm$^{-1}$. 

Figure 2 schematically illustrates the relative orientations of the molecular beam, the laser beam, the laser polarization ($\vec{F}_{\rm{L}}$), the axis of the cylindrically symmetric electrode stack used to 
generate the electric field ($\vec{F}_{\rm{dc}}$) and the axes of the two solenoid pairs used to produce the magnetic fields ($\vec{B}_{\rm{par}}$ and $\vec{B}_{\rm{perp}}$). The atomic 
beam propagates in a direction parallel to the electric field. The magnetic field can be produced at arbitrary angle
to the electric field in the $y'$, $z'$ plane 
by adjusting the currents flowing through the two pairs of coils. When the magnetic-field vector points parallel to the He* beam and the He$^+$ flight axis ($z'$ axis in figure 2), the Lorentz
force does not deflect the ions. Consequently, the maximal value of the magnetic field in $z'$ direction is only limited by the current that can be applied to the coils, i.e., 200 A, 
corresponding to $B_{\rm{par}} = 280$ mT. When the magnetic field is applied in the direction perpendicular to the ion flight axis, the Lorentz force deflects the ions, which makes it impossible to 
record spectra for $B_{\rm{perp}} > 20$ mT.   

The laser polarization used for the experiments is chosen so as to be predominantly 
perpendicular to both the electric and magnetic fields, implying the selection rule $\Delta m_l = \pm1$ for the transition from the 2 $^3$S$_1$ state to the Rydberg states. 
In some experiments, the laser polarization was slightly tilted (less than 5$\degree$), 
so that the $m_l$ = 0 component in the pure Zeeman spectra and the 31s level in the pure Stark spectra and the spectra recorded for parallel fields could also be observed. 
The effects of the slightly tilted polarization on the spectra recorded in combined electric and magnetic fields are too 
weak to be observed. The calibration of the magnetic field is performed by measuring the Zeeman splitting 
for a range of currents applied to the two pairs of coils and determining the relationship between current and magnetic field for each pair of coils separately. 
Since the beam velocity is parallel to the applied electric field, the motional Zeeman effect is zero. The motional 
Stark effect resulting from the magnetic-field component perpendicular to the beam propagation axis is 0.09 V/cm for $B$ = 8 mT and is negligible compared to the typical
electric-field strengths used in this work.     

\subsection{Theoretical treatment}
The Hamiltonian describing a He Rydberg atom in combined electric and magnetic fields with a relative orientation specified by the angle $\alpha$ is given in atomic units by 

\begin{eqnarray}
\fl \hat H &= \hat H_0 + {1 \over 2}B(\hat l_z-g_e \hat S_z) + {1 \over 8}B^2(\hat x^2 + \hat y^2) \nonumber\\ 
        & + F \hat z \cos \alpha + F \hat x \sin \alpha, 
\end{eqnarray}
if the $z$ axis is chosen to coincide with the magnetic field. The magnetic field is given in atomic units of $\hbar/(ea_0^2)$ = 2.35$\times 10^5$ T and the electric field in 
atomic units of $E_h/(ea_0)$ = 5.14 $\times 10^9$ V/cm. $\hat H_0$ is the unperturbed (zero-field) Hamiltonian of the helium atom. The zero-field energies are calculated from Rydberg's formula using the known quantum defects 
for triplet ($S$ = 1) helium ($\delta_s$ = 0.2967, $\delta_p$ = 0.0684, $\delta_d$ = 0.0029 and $\delta_{f, g, ...}$ = 0 \cite{martin87a}). The second term in the Hamiltonian (1) is the 
paramagnetic term and $\hat l_z$ and $\hat S_z$ are the orbital and spin angular-momentum operators of the Rydberg electron in the direction of 
the magnetic field ($z$ axis), respectively, and $g_e \approx -2.00231$ is the electron $g$ factor. The diamagnetic term ${B^2 \over 8}(\hat x^2 + \hat y^2)$  
 can be neglected if $n^4B << 1$ and does not play a significant role at $n$ = 30 for the magnetic field strengths used in the current study, except for 
the spectra recorded at magnetic field strengths above 100 mT. 
The last two terms in the Hamiltonian (1) describe the interaction with the electric field along the $z$ and $x$ axes.

If an electric or a magnetic field is applied, the orbital angular momentum quantum number $l$ is no longer a good quantum number. 
Only the projection of the orbital angular momentum onto the axis of the field, $l_z = m_l\hbar$, with $m_l$ = -$l$, -$l$+1, ..., $l$, is a constant of motion. The same is true for parallel electric 
and magnetic fields. When these fields are not parallel, also the cylindrical symmetry is broken and not even $m_l$ is a good quantum number any more. 

The diamagnetic term is diagonal in $m_l$ but mixes all $l$ states of the same parity according to the selection rule  $\Delta l = 0, \pm 2$ without restriction on $\Delta n$.  
The Stark Hamiltonian (last two terms in Hamiltonian (1)) has components along the $x$ and $z$ axes and couples states according to the selection rule $\Delta l = \pm 1$.
The $z$ component conserves the quantum number $m_l$, whereas the $x$ component 
mixes states differing in $m_l$ by $\pm 1$. Both components can couple states of different $n$ values, but at the electric fields used in this investigation 
the coupling between adjacent $n$ manifolds is  small compared to the coupling within one $n$ manifold. 
The paramagnetic term is diagonal in $n$, $l$ and $m_l$. 

In the cases of a strong magnetic field or of a very weak spin-orbit coupling, $\hat S$ and $\hat l$ couple more strongly to the magnetic field than to 
each other and precess independently about the magnetic-field direction. The total angular momentum $\hat J =\hat l + \hat S$ of the Rydberg electron is 
no longer a constant of motion, but $\hat J_z =\hat l_z + \hat S_z$ is. This situation corresponds to the Paschen-Back regime, which describes the $n$ = 30 and 45 Rydberg 
states of He explored in this work accurately, 
because the spin-orbit interaction is very weak in a light atom such as He and scales as $\sim{1 \over n^3}$. The Paschen-Back regime and the selection 
rules $\Delta S = 0$, $\Delta m_S = 0$ allow us to ignore the spin part of the Hamiltonian. All spectra presented in this article were recorded in the regime where the magnetic interaction is much 
weaker than the Coulomb interaction but at the same time much stronger than the spin-orbit interaction. The spectra for magnetic-field strengths beyond 100 mT were calculated using Hamiltonian (1) 
but disregarding the electron spin.   
The Hamiltonian can also be expressed in a coordinate system with the $z$ axis chosen to coincide with the electric field  

\begin{equation}
\left.
 \hat H = \hat H_0 + F \hat z + {1 \over 2}B\hat l_z \cos \alpha + {1 \over 2}B\hat l_x \sin \alpha. 
\right.
\end{equation}  
This coordinate system is not suitable for expressing the diamagnetic interaction, and this Hamiltonian was used only for weak magnetic fields ($B <$ 20 mT).  
In this case, the paramagnetic term along the $x$ axis is diagonal in $l$ and $n$ but mixes states with $\Delta m_l = \pm 1$. To compute the 
Rydberg-excitation spectra, the Hamiltonians (1) or (2) are expressed in matrix form using $|nlm_l\rangle$ = $R_{nl}(r)Y_{lm_l}(\theta, \phi)$ 
basis functions and the line positions and intensities are derived from its eigenvalues and eigenfunctions. 

The number of states that need to be included in the calculations to accurately reproduce the 
level positions depends on the electric- and magnetic-field strengths \cite{lahaye89a}. For the electric- and magnetic-field strengths used in the present investigation 
to record spectra around $n$ = 30 ($F \leq 78$ V/cm and $B \leq 280 $ mT),
the basis set had to include all states with $n$ = 29 - 32, $l$ = 0, ..., $n$-1 and $m_l$ = -$l$, -$l$+1, ..., $l$, i.e., a total of 3726 functions, to correctly 
reproduce the line positions. Using Hamiltonian (2) instead of (1) enables one to reduce the size of the basis set and to include only states with $|m_l| \leq 2$ in calculations 
performed for large electric fields  
and small magnetic fields. No significant difference was found in the spectra calculated for perpendicular fields of 78 V/cm and $B <$ 20 mT 
with either a full $|m_l|\leq l$ basis set or the reduced $|m_l| \leq 2$ basis set.  The radial part of the matrix elements was computed using $R_{nl}(r)$ radial functions evaluated numerically for $l$ = 0 - 2 using 
Numerov's integration method in combination with the known values of the quantum defects, following the procedure described by Zimmerman \it et al. \rm  for alkali-metal atoms \cite {zimmerman79a}.     
The spectra for magnetic-field strengths below 20 mT were calculated with the Hamiltonian (2) including the full $m_l$ = -$l$, ..., $l$ basis set. Calculations 
based on Hamiltonian (2) with the truncated basis of $m_l$ states were exploited in the calculation of correlation diagrams, when the use of a full $m_l$ basis made it more 
difficult to assign quantum numbers to a given state.         

Single-photon excitation from the metastable 2 $^3$S$_1$ level with radiation polarized linearly in the $x',y'$ plane provides access to the $l$ = 1, $m_l$ = $\pm$1 
components of the Rydberg wave functions. Consequently, the relative spectral intensity $I_i$ of a transition to the Rydberg states $i$ mixed by the Stark and 
Zeeman effects can be determined as 

\begin{equation}
\left.
 I_i = \left|\sum_n c_{n11}^{(i)}+ c_{n1-1}^{(i)} \right|^2, 
\right.
\end{equation}  
where $c_{n11}^{(i)}$ and $c_{n1-1}^{(i)}$ represent the coefficients of the $|n11\rangle$ and $|n1-1\rangle$ basis functions in the $i$-th eigenfunction. 

\section{Results and discussion}
\subsection{Spectra recorded in pure magnetic fields}
Figure 3(a) presents the energy-level structure calculated for $n$ = 30, $m_l = \pm1$ levels of He for 
magnetic fields in the range 0 - 250 mT. The main effect of the field is to split 
each zero-field level into the two $m_l = \pm1$ components separated by $2\mu_BB$.  
The effects of the diamagnetic term of the Hamiltonian become apparent at fields beyond $\sim$100 mT as (1) a splitting of the high-$l$ 
manifold ($l > 2$) of Zeeman levels into 27 levels for each of the $m_l$ = 1 and $m_l$ = -1 group, (2) the gradual integration of the d Zeeman 
levels in the high-$l$ manifold, and (3) an asymmetric splitting of the p $m_l = \pm1$ levels with respect to the zero-field position of the p state. 

For magnetic fields below 230 mT, the $l$ mixing of the optically accessible 30p level with nonpenetrating $l$ = 3, 5, 7, ..., 29 levels induced 
by the diamagnetic term of the Hamiltonian [see Hamiltonian (1)] is still extremely weak, so that only the 30p level could be observed 
within the sensitivity limit of our experiments. Calculations based on formula (3) indicate that the intensities of transitions 
to the high-$l$ manifold of $n$ = 30 Zeeman levels are more than 1000 times weaker than the transitions to the p state and that these transitions would only become observable at magnetic 
fields of $\sim$750 mT, which are, unfortunately, not accessible with our solenoids. Consequently, only the third effect of the diamagnetic 
term listed above can be observed experimentally at $n$ = 30. 

Figures 3(b) and (c) depict the Zeeman spectra recorded in the vicinity of the 2 $^3$S $\rightarrow$ 30 $^3$P transition for a perpendicular polarization 
of the UV laser [figure 3(c)] and for a slightly tilted polarization [figure 3(b)] and compares the spectra with the spectra calculated 
for $\Delta m_l = \pm1$. The observation of the $m_l$ = 0 component in figure 3(b) makes it possible to recognize the asymmetric 
splitting of the $m_l = \pm1$ levels.   

\begin{figure}
 \includegraphics{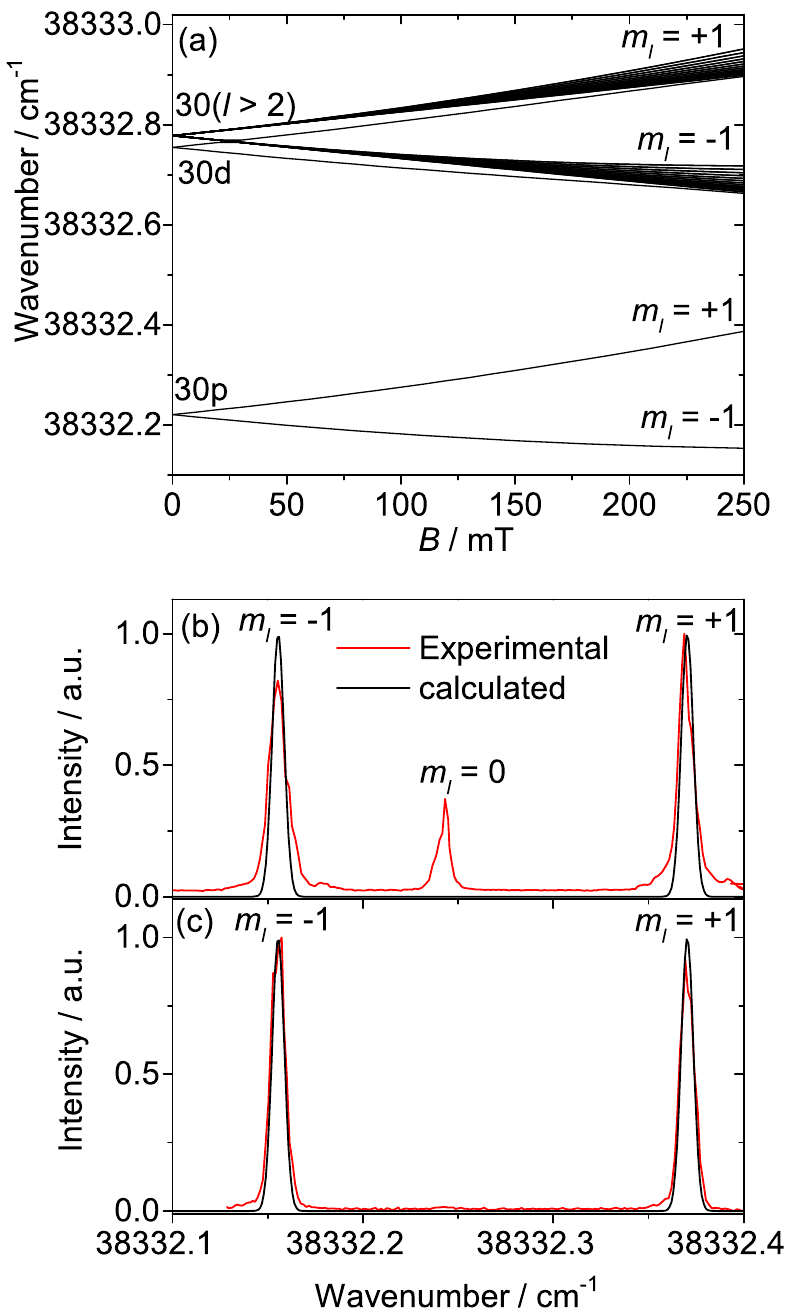}
 \caption{\label{} (a) Magnetic-field dependence of the energy-level structure of $n = 30,\ m_l = \pm1$ Rydberg states of helium. 
Experimental and calculated spectra of transitions from the 2 $^3$S$_1$ state of He to the 30 $^3$P, $m_l = \pm1$ states recorded in a magnetic field $B$ = 230 mT with the laser polarization vector perpendicular to the magnetic-field vector (c), and slightly tilted away from the perpendicular arrangement (b).}
 \end{figure} 

To verify that our numerical procedure to calculate Zeeman spectra correctly accounts for the diamagnetic term, spectra were also measured for $n$ = 45 at a field of 272 mT.
At this $n$ value and this magnetic-field strength, the diamagnetic interaction, which scales as $\sim n^4$, is strong enough to induce significant $l$ mixing and to make all Zeeman levels of negative 
parity observable, as illustrated by figure 4. The pure Zeeman spectra follow the general trends of the diamagnetic Zeeman effect discussed in previous  
articles \cite{veldt92a,holle86a,wintgen86a} and are labeled with the state index $K$, which, at $n$ = 45, ranges from 0 to 42 in steps of 2 \cite{veldt92a}. 
A dashed vertical line, called the separatrix in earlier work \cite{veldt92a,holle86a,wintgen86a}, divides the Zeeman levels into two groups regarded as "vibrational" and "rotational" states \cite{veldt92a}.    

\begin{figure}
 \includegraphics{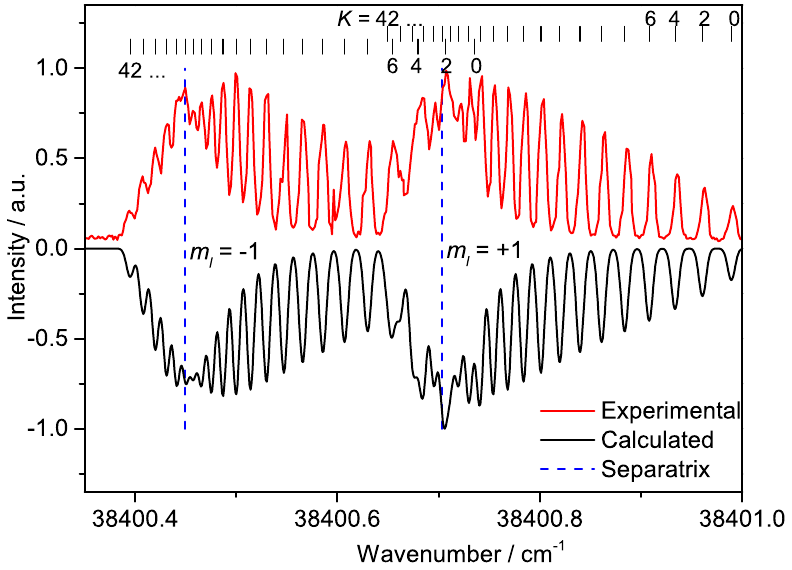}
 \caption{\label{} Experimental and calculated spectra of the $\Delta m_l = \pm1$ transitions from the 2 $^3$S$_1$ state of helium to the Zeeman levels of 
the $n$ = 45 manifold at a magnetic field $B$ = 272 mT.}
 \end{figure} 

\subsection{Spectra recorded in pure electric fields}

\begin{figure}
 \includegraphics{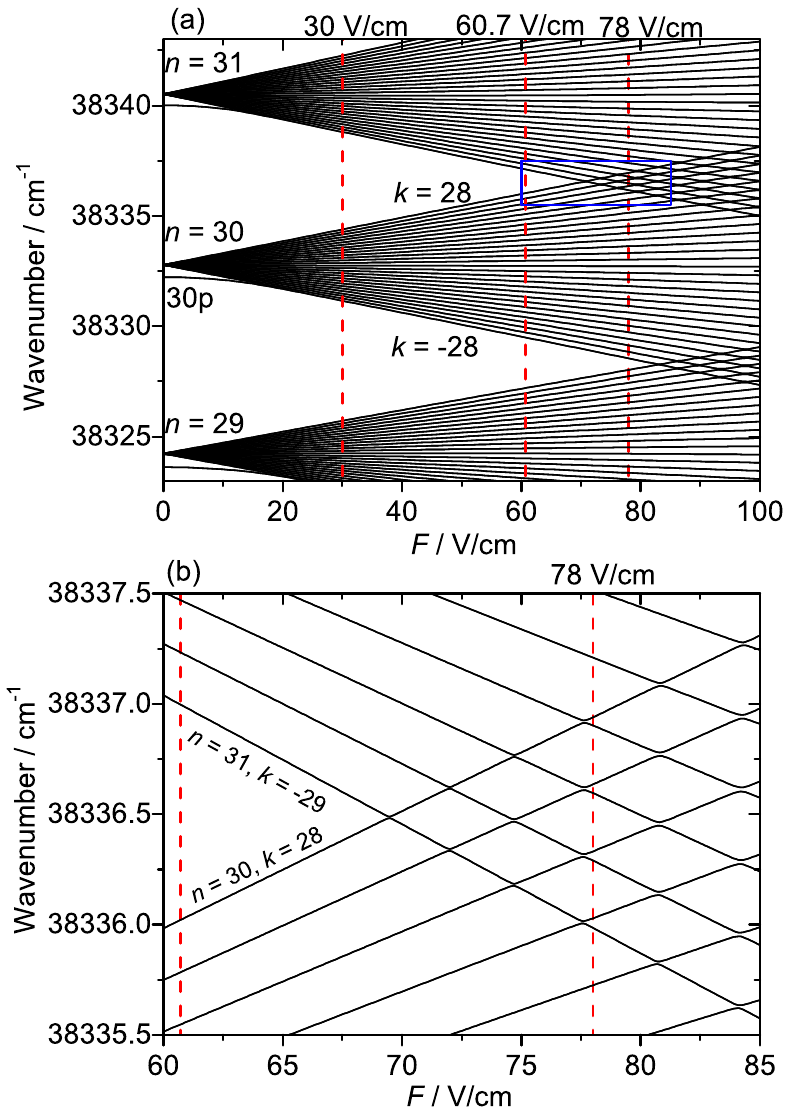}
 \caption{\label{} (a) Electric-field dependence of the $|m_l|$ = 1 energy levels of Rydberg helium ($n$ = 29 - 31). The electric-field strengths of 30, 60.7 and 78 V/cm at which the Stark spectra were measured 
are indicated as vertical red dashed lines. (b) Magnified view of the region of avoided level crossings marked by a blue rectangle in panel (a).}
 \end{figure}

 Figure 5 presents the results of calculations of the Stark effect of ($S$ = 1) He Rydberg states with $n$ in the range 29 - 31. The calculations were 
carried out starting from Hamiltonian (2) with $B$ = 0 T using only $m_l = \pm1$ levels in the basis, which rules out any contribution from $n$s levels. 
The quantum defect of the $n$p series is large enough so that the p levels appear well separated from the manifold of high-$l$ ($l \geq 2$) Stark levels and 
are subject to a quadratic Stark shift at low fields. The field at which neighboring Stark manifolds start overlapping  - this field is known as 
Inglis-Teller field \cite{gallagher94a,Inglis39a} and can be calculated in atomic units as $F_{\rm{IT}} = {1 \over 3n^5}$ - is 71 V/cm between $n$ = 30 and 31 and 83 V/cm between $n$ = 29 and 30.

\begin{figure}
 \includegraphics{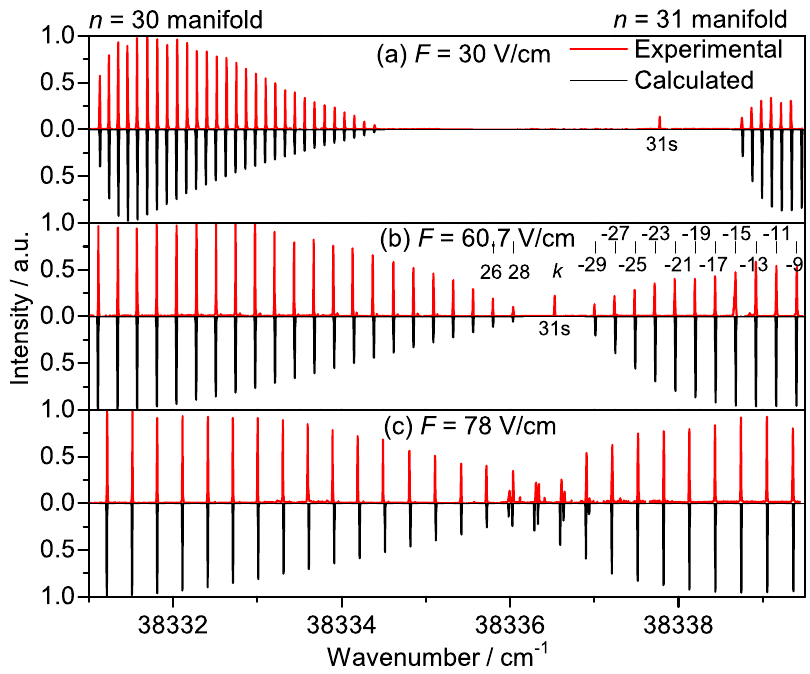}
 \caption{\label{}Experimental and calculated Stark spectra of Rydberg helium for the electric-field strengths (a) $F$ = 30 V/cm, 
                 (b) $F$ = 60.7 V/cm and (c) $F$ = 78 V/cm.}
\end{figure}

The experimental Stark spectra recorded at fields of 30, 60.7 and 78 V/cm (indicated by vertical dashed lines in figure 5) are compared with calculated spectra in figure 6. Apart from 
the line corresponding to the 31s level (this line is present in the experimental spectra because of the slight tilt of the UV laser polarization vector away from a 
perfect perpendicular arrangement (see Section 2.1), but absent from the calculated spectra), the agreement between experimental and calculated spectra is excellent. 
The spectra recorded for $n$ = 30 consist of 29 equally spaced lines (not all shown in figure 6) corresponding to the $|m_l| = 1$ Stark states with Stark index $k$ running 
from -28 to 28 in steps of two. At 78 V/cm [figure 6(c)], the $n$ = 30 and 31 Stark manifolds partially overlap, which leads to a more congested spectrum in 
the region of overlap. The crossings between the Stark states of different $n$ manifolds are weakly avoided. The avoided crossings are difficult to see in figures 5(a) and 6(c), but can 
be seen on the enlarged scale of figure 5(b).

\subsection{Spectra recorded in parallel electric and magnetic fields}
\begin{figure}
 \includegraphics{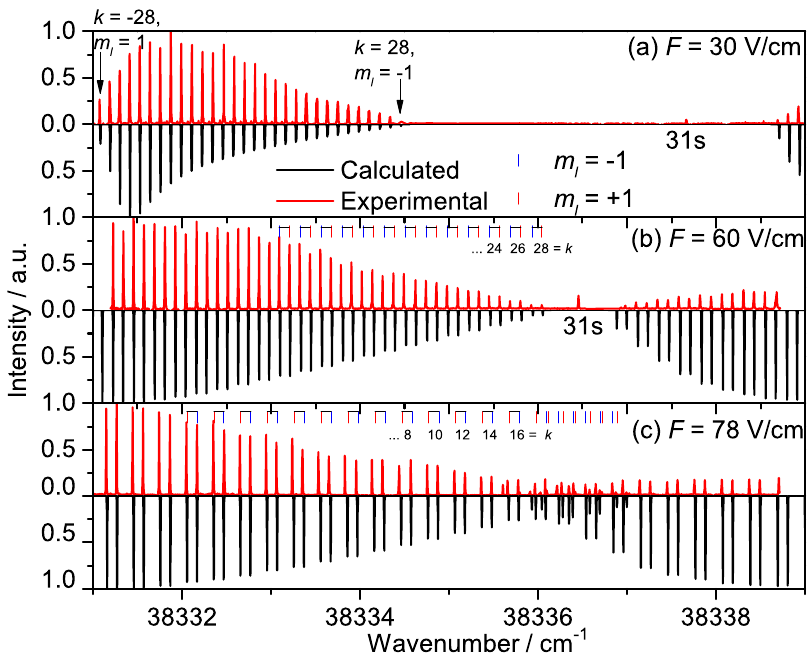}
 \caption{\label{}Experimental and calculated spectra of the transitions from metastable He to the Rydberg states of He near $n$ = 30 for parallel magnetic ($B$ = 120 mT) 
and electric [(a) $F$ = 30 V/cm, (b) $F$ = 60 V/cm and (c) $F$ = 78 V/cm] fields.}
 \end{figure}
 
The case of Rydberg atoms in weak parallel electric and magnetic fields has been extensively studied in hydrogen \cite{cacciani92a}, lithium \cite{cacciani88b,cacciani88a}, 
helium \cite{veldt93a} and theoretically in alkali-metal atoms \cite{santos99a}.
The results obtained in the present study for the $S$ = 1 states of He are briefly summarized here for completeness. The emphasis is placed on the two special cases where the diamagnetic 
term has only minor effects and both paramagnetic Zeeman and Stark effects are linear  ($n$ = 30, $B$ = 120 mT and $F$ = 30 - 78 V/cm), and where the diamagnetic term is dominant and the Stark effect is 
linear ($n$ = 45, $B$ = 277 mT and $F$ = 0.7 - 8 V/cm). 

A comparison between experimental and calculated spectra in the vicinity of the $n$ = 30 manifold for $B$ = 120 mT and $F$ = 30, 60 and 78 V/cm is presented in figure 7. Figure 8(a) provides an 
overview of the calculated energy level structure for $B$ = 120 mT and $F$ in the range 0 - 78 V/cm (blue lines) and also displays sections of the 
spectra (black lines) calculated at $F$ = 30, 60 and 78 V/cm.      
At 60 and 78 V/cm, the effect of the magnetic field is to split the $m_l = \pm1$ components of all Stark states in doublets separated by $2\mu_BB$. At 60 V/cm, the splitting 
is almost exactly half the spacing between adjacent Stark states so that the spectrum appears as a single regular series of lines. At 30 V/cm, the Stark manifold consists of 
30 lines instead of the 29 lines expected for a $|m_l| = 1$, $n$ = 30 Stark manifold. The reason for the additional line is the fact that the Zeeman doublets have the same spacing 
as the adjacent members of the Stark manifold. Consequently, each of the inner 28 lines corresponds to two transitions, one to the $k$, $m_l$ = -1 state and the other to the $k+2$, $m_l$ = 1 state. 

\begin{figure*}
\includegraphics{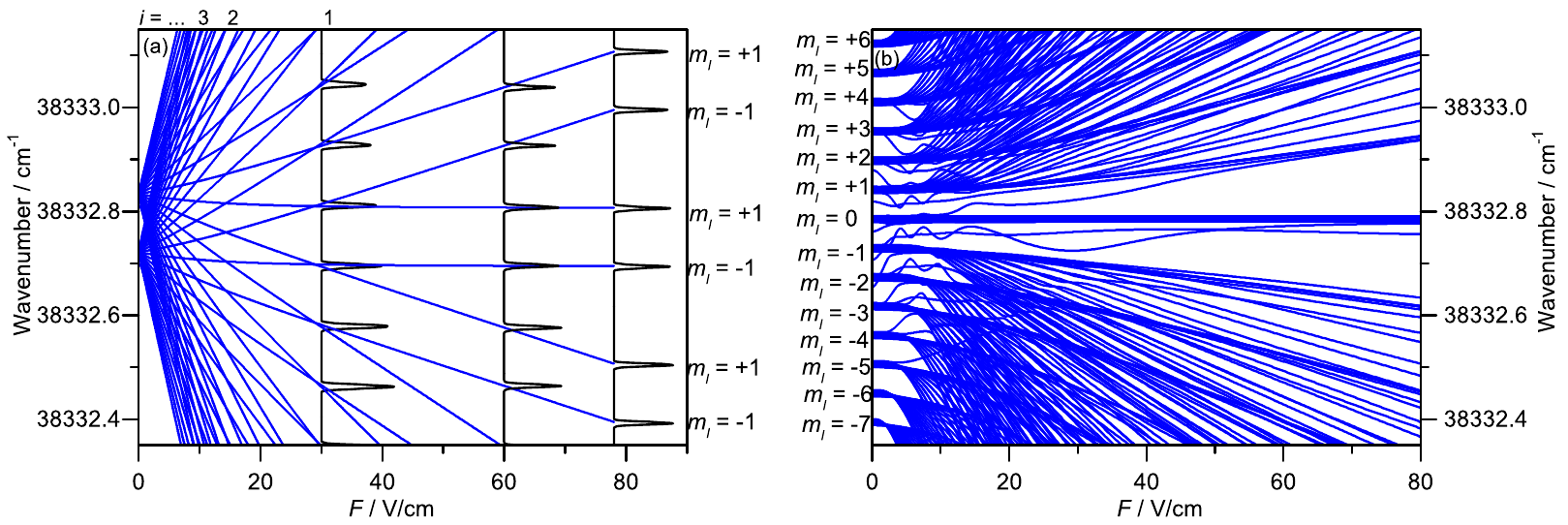} 
\caption{\label{} Electric-field dependence of the $n$ = 30, $S$ = 1 Rydberg states of helium calculated for a magnetic field $B$ = 120 mT and  
(a) parallel ($\alpha = 0\degree$) and (b) perpendicular ($\alpha = 90\degree$) arrangements of the fields. The spectra 
for $B$ = 120 mT, $\alpha = 0\degree$ and $F$ = 30, 60 and 78 V/cm are also depicted in panel (a) to illustrate the structure of the spectra presented in figure 7.}
\end{figure*}

The approximate linearity of both Zeeman and Stark effects implies the existence of $n-|m_l|-1$ sets of crossings between states of the $m_l$ = +1 and $m_l$ = -1 manifolds at electric 
fields approximately given by the condition (in atomic units)    

\begin{equation}
\left.
 F = {B \over 3ni};  (i = 1, 2, ..., n-|m_l|-1). 
\right.
\end{equation}  
The region where these intermanifold crossings take place is limited to the electric field range between 1.04 V/cm ($i$ = 28) and 29.2 V/cm ($i$ = 1). Not all 
levels undergo all 28 crossings: The $k$ = 28, $m_l$ = +1 level does not cross any other level, the $k$ = 26, $m_l$ = +1 level crosses only one level ($k$ = 28, $m_l$ = -1) 
and only the $k$ = -28, $m_l$ = +1 level undergoes all 28 crossings. The exact nature of the crossings, which results from the fact that $m_l$ is a good quantum number when the 
fields are parallel, render the appearance of the level structure simple in the range where both Zeeman and Stark effects are linear (see figure 8(a)). The situation changes 
completely when the fields are perpendicular, as illustrated by the level structure presented for comparison in figure 8(b) (see also section 3.4). The coupling of states of 
different $m_l$ values leads to many avoided crossings and characteristic groups of levels, merging at low fields to states of well-defined $m_l$ values.    

Figure 9 compares experimental and calculated spectra of $n$ = 45 Rydberg states of helium for parallel fields $B$ = 277 mT and $F$ = 0.7, 3 and 7.9 V/cm, i.e., in a regime where the 
diamagnetic interaction is significant. The agreement between experimental (red lines) and calculated (black lines) is excellent for all three electric-field strengths. To characterize the 
spectra recorded in the regime of combined Stark and diamagnetic effects, the approximate constant of 
motion $\Lambda_{\beta}$ was exploited, given by \cite{braun84a,cacciani88a,cacciani88b,cacciani88c} 
\begin{equation}
\left.
 \Lambda_{\beta} = 4A^2-5A^2_z+10\beta A_z, 
\right.
\end{equation} 
where $A$ is the Runge-Lenz vector, $A_z$ is its projection along the field axis, and the parameter 
\begin{equation}
\left.
 \beta = {12F \over 5n^2B^2}, 
\right.
\end{equation} 
represents the relative strength of the linear Stark interaction with respect to the diamagnetic one. The eigenvalues of Hamiltonian (1) (neglecting electron spin) can be 
written as $E = E_0 + E_{\rm{p}} + E_{\rm{ds}}$, 
where $E_0$ is the zero-field energy, $E_{\rm{p}}$ is the paramagnetic shift and  
\begin{equation}
\left.
 E_{\rm{ds}} = {1 \over 16}B^2n^2(n^2+m_l^2+n^2\Lambda_{\beta}) 
\right.
\end{equation}
describes the contribution of the diamagnetic interaction and the linear Stark effect. 

\begin{figure}
 \includegraphics{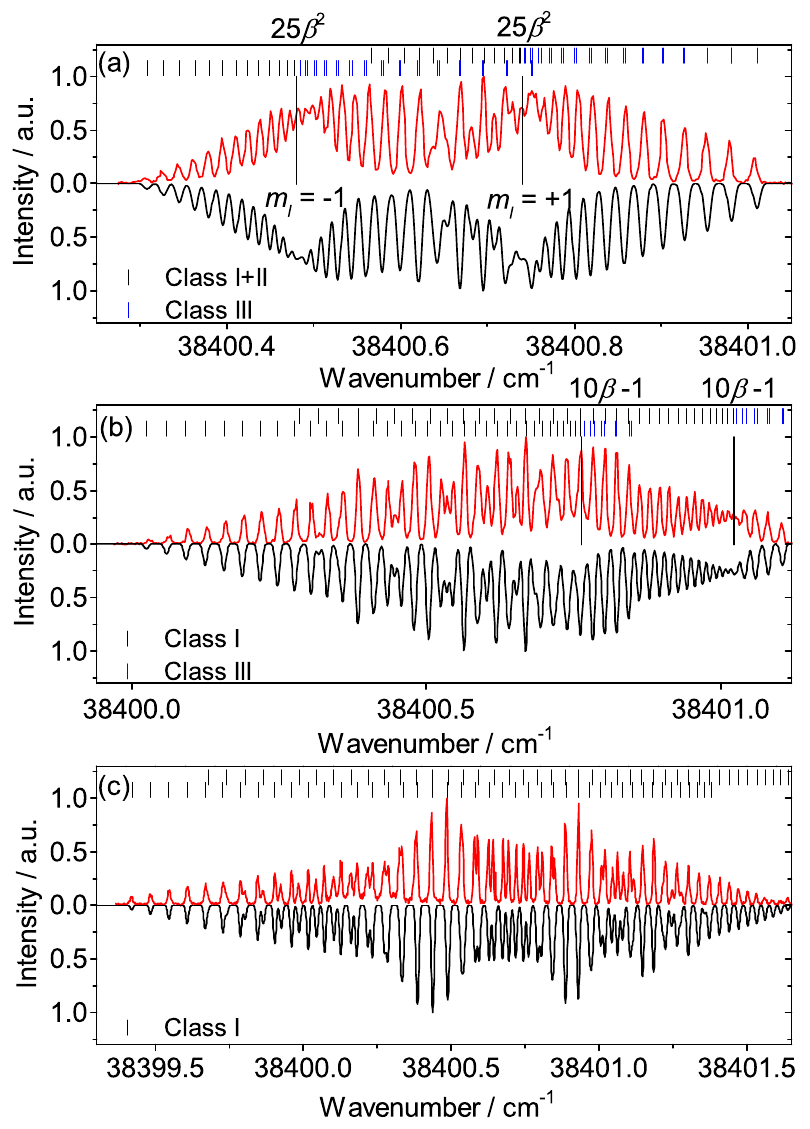}
 \caption{\label{} Experimental (red) and calculated (black) spectra of transitions from the 2 $^3$S$_1$ state of 
He to $n$ = 45, $m_l = \pm1$ Rydberg states recorded at a magnetic field of 277 mT and electric fields of (a) 0.7 V/cm, (b) 3 V/cm and (c) 7.9 V/cm in a parallel arrangement of the field vectors. The positions corresponding to $\Lambda_\beta = 25\beta^2$ 
and $\Lambda_\beta = 10\beta - 1$ are shown in panel (a) and (b), respectively, as vertical solid lines.}
\end{figure} 

In the case of parallel electric and magnetic fields, three classes of states can be observed \cite{cacciani88a,cacciani88b,cacciani88c,veldt93a}: vibrational states with positive 
(class I) or negative (class II) dipole moment and 
rotational states (class III). Vibrational and rotational states are separated from each other by a separatrix as 
already discussed in Section 3.1. The position of the separatrix is given by $\Lambda_{\beta} = 25\beta^2$ \cite{braun84a}. Class III states 
transform one by one into class I states when $\Lambda_{\beta}$ approaches $25\beta^2$ or $10\beta-1$ \cite{braun84a}.    
The field dependence of the $m_l$=1 level structure at $n$ = 45 and $B$ = 277 mT is depicted in figure 10, which  
reveals a linear Stark effect for both types of vibrational states and a quadratic Stark effect for the rotational states. The interaction of 
the electric field with the dipole moment induced by the magnetic field results in positive (negative) energy shifts for class II (class I) states. Consequently, a multitude of level 
crossings occur between states of classes I and II at low fields.  

The values of $\beta$ and $\Lambda_{\beta}$ determine the ranges in which the different classes of states exist \cite{cacciani88a,cacciani88b,cacciani88c,veldt93a}. At $\beta$ values 
below ${1\over 5}$, all three classes of states coexist. Classes I and II correspond to the lowest $\Lambda_{\beta}$ values and are encountered in the ranges ($-1-10\beta \leq \Lambda_\beta \leq 25\beta^2$) 
and ($-1+10\beta \leq \Lambda_\beta \leq 25\beta^2$), respectively, whereas class III states are found in the range ($25\beta^2 \leq \Lambda_\beta \leq 4+5\beta^2$). The spectrum depicted 
in figure 9(a), recorded at $F$ = 0.7 V/cm and $B$ = 277 mT ($\beta = 0.12$) corresponds to this situation, and nicely reveals the structure expected for vibrational (black assignment marks) 
and rotational (blue assignment marks) states.

\begin{figure}
 \includegraphics{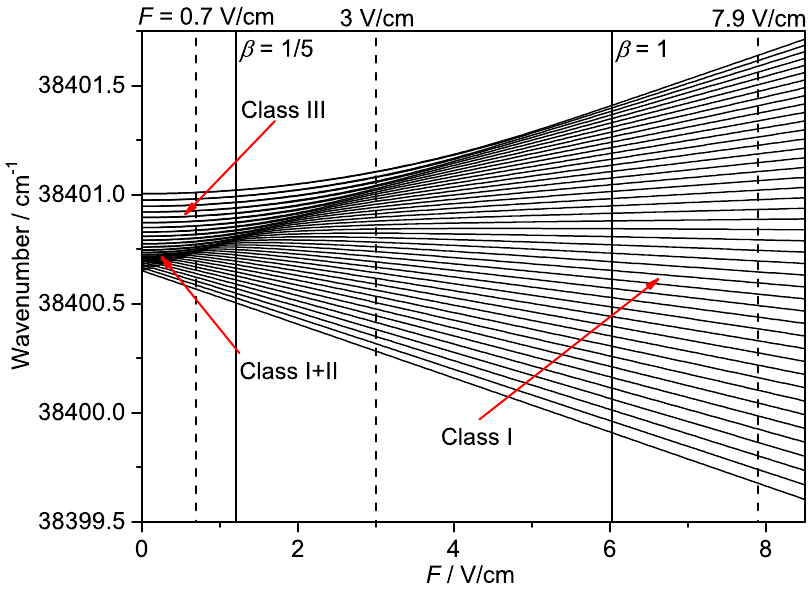} 
 \caption{\label{}Electric-field dependence of $n$ = 45, $m_l = +1$ Rydberg state of He calculated for a fixed magnetic field $B$ = 277 mT and $\alpha = 0\degree$.  
Vertical dashed lines indicate the electric-field strengths $F$ = 0.7 V/cm, 3 V/cm and 7.9 V/cm at which the spectra presented in figure 9 were measured. Solid lines divide 
the map into three regions with $\beta \leq {1 \over 5}$, ${1 \over 5} < \beta \leq 1$ and $\beta > 1$.}
 \end{figure}

States belonging to class II with negative dipole moments are no longer encountered beyond $\beta = {1\over 5}$. Between $\beta = {1\over 5}$ and $\beta = 1$ class I and III states 
coexist and have $\Lambda_{\beta}$ values in the ranges ($-1-10\beta \leq \Lambda_\beta \leq -1+10\beta$) and 
($-1+10\beta \leq \Lambda_\beta \leq 4+5\beta^2$), respectively. The spectrum recorded at $F$ = 3 V/cm and $B$ = 277 mT ($\beta = 0.5$) and depicted in figure 9(b) corresponds to this situation. 
As $\beta$ increases, class III states are gradually converted into class I states. At $\beta = 0.5$, the majority of states already belong to class I and form long anharmonic 
progressions ending at the positions of the two separatrices marked by vertical lines in figure 9(b). For $\beta > 1$, only class I states ($-1-10\beta \leq \Lambda_\beta \leq -1+10\beta$) exist.
This situation is illustrated by the spectrum displayed in figure 9(c) which was recorded at $F$ = 7.9 V/cm and $B$ = 277 mT ($\beta = 1.33$) and also by the spectra presented in figure 7, for which              
$\beta \gg 1$. This situation can be described by two progressions of almost equidistant Stark states, one with $m_l = +1$ and the other with $m_l = -1$, separated by $2\mu_BB$.  

\subsection{Spectra recorded in perpendicular electric and magnetic fields}

\begin{figure*}
 \includegraphics{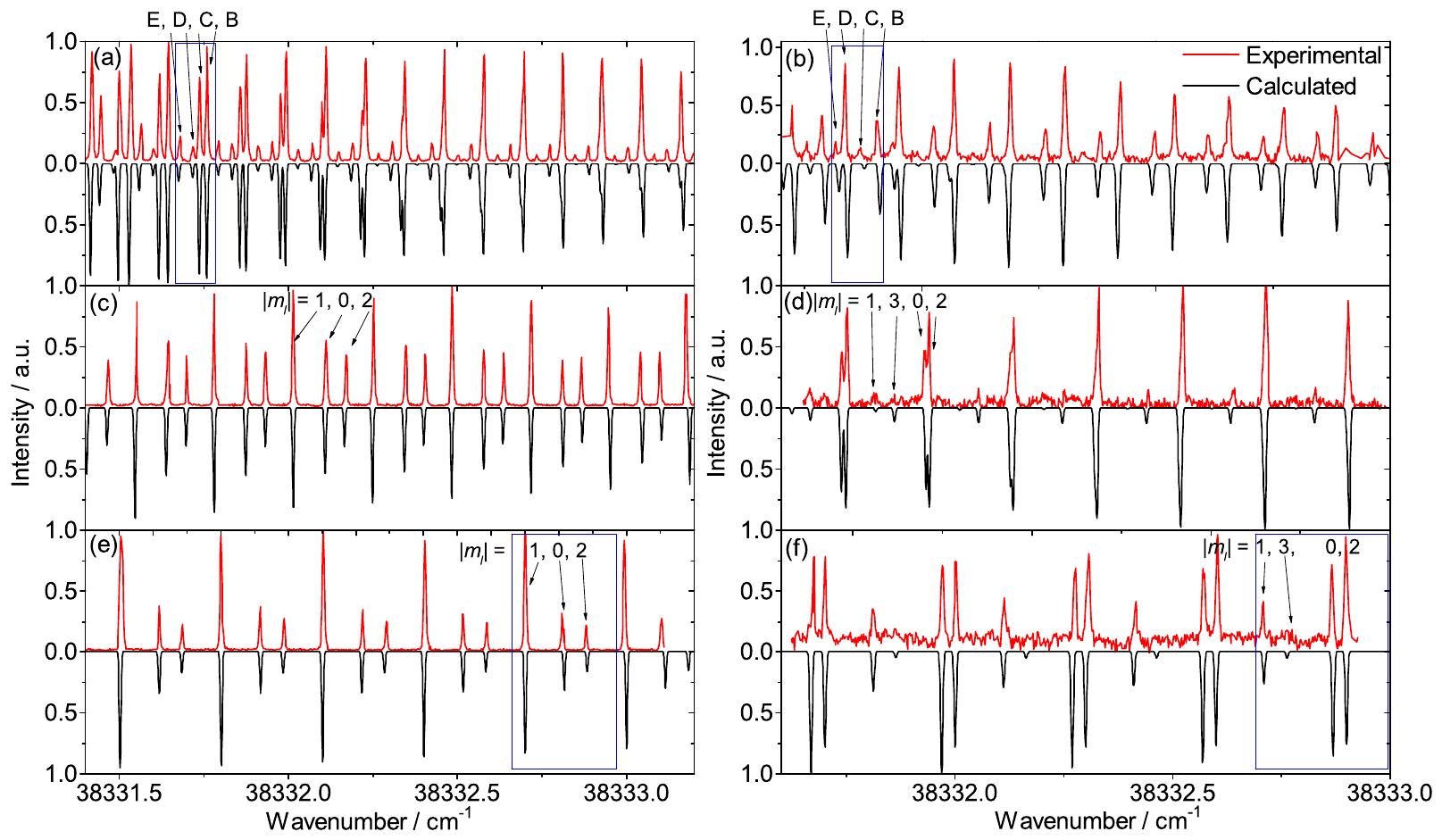}
 \caption{\label{}Experimental and calculated spectra of $n$ = 30 He Rydberg states for perpendicular electric [(a-b) $F$ = 30 V/cm, (c-d) $F$ = 60 V/cm and (e-f) $F$ = 78 V/cm] and 
magnetic [$B$ = 7.2 mT (left) and 15 mT (right)] fields. The parts of the spectra enclosed in the blue frames are shown on an enlarged scale in figures 12 and 13.}
 \end{figure*}

The experimental and calculated spectra for perpendicular electric and magnetic fields are presented in figure 11 for magnetic fields of 7.2 mT and 15 mT, 
respectively, and for electric fields $F$ = 30, 60 and 78 V/cm. As explained in Section 2.1, the deflection of the He$^+$ ions away from the 
detection axis by the Lorentz force prevents the use of perpendicular magnetic fields larger than 20 mT and reduces the signal-to-noise ratio of the spectra recorded at 15 mT shown in panels (b), (d) and (f). The calculated spectra reproduce all features of the experimental spectra, 
which consist of regularly spaced groups of three to five levels. The spacing between these level groups corresponds to the linear Stark effect. 
Although $m_l$ is not a good quantum number in this situation, approximate spectral assignment in terms of $|m_l|$ can be 
performed by exploiting the adiabatic correlations to the situation of zero-magnetic field, where $m_l$ is a good quantum number and levels differing in the sign of $m_l$ are degenerate.  

Figure 12(a) shows a selected region of the calculated map of levels in dependence of the perpendicular magnetic-field strength at an electric-field strength of $F$ = 78 V/cm. 
The figure depicts the eigenvalues of the Hamiltonian (2) set up with a basis limited to $m_l = -3, ..., +3$ levels. Also shown in the figure are the 
calculated spectra for $B$ = 7.2 and 15 mT, which correspond to the regions enclosed in blue frames in figures 11(e,f). At these magnetic-field strengths, the spectra calculated by including 
only $m_l = -3, ..., +3$ are almost identical to the spectra calculated using the full $m_l = -l, ..., +l$ basis, indicating 
a negligible contribution of levels with $|m_l| > 3$.  

The spectrum measured at $B$ = 7.2 mT and $F$ = 78 V/cm exhibits repeated structures of three peaks (see region enclosed in a blue frame in figure 11(e) and displayed on an enlarged 
scale in figure 12(a)). 
The strong transitions are to levels correlating adiabatically to a 
zero-magnetic-field level with $|m_l|$ = 1, whereas the two weaker transitions are to levels correlating to $|m_l|$ = 0 and degenerate $m_l = \pm2$ levels at $B$ = 0 mT. 
The same labels can be assigned to the peaks in the spectrum recorded at $F$ = 60 V/cm and $B$ = 7.2 mT (figure 11(c)), but the intensities of the lines linked 
to the $|m_l|$ = 1 levels are weaker and those of the lines correlating to $|m_l| = 0$ and $|m_l| = 2$ levels are stronger, than at 78 V/cm. At a magnetic field 
of 15 mT (figure 11(f) and 12(a)), the intensity of 
the $|m_l|$ = 1 line is reduced and its position shifts to lower energies, whereas the $|m_l|$ = 0 and $|m_l| = 2$ levels get closer to each other and 
gain intensity. The energy levels with $|m_l| = 2$ are not degenerate any more at a magnetic field of 15 mT and only the transition to the lower level 
has nonzero intensity. A line corresponding to a transition to a $|m_l| = 3$ level is visible in the calculated spectrum but is too weak to be observed experimentally. 
The transitions to the $|m_l|$ = 0 and $|m_l| = 2$ levels are hardly distinguishable and even merge into a single line in the higher 
energy part of the $F$ = 60 V/cm spectrum (figure 11(d)) and an additional peak appears associated to a level correlating to $|m_l| = 3$ at $B$ = 0 mT.

The degree of $m_l$ mixing in a given state $i$ can be quantified by evaluating the sum     
 \begin{equation}
\left.
 p^{(i)}(m_l) = \sum_{n,l} |c_{n,l,m_l}^{(i)}|^2
\right.
\end{equation}  
for each $m_l$ value.
The distributions of $p^{(i)}(m_l)$ values for the final states observed in the spectra recorded at 7.2 and 15 mT and 78 V/cm are presented in figures 12(c) and (d), respectively. 
The dominant $|m_l|$ character of the $p^{(i)}(m_l)$ distribution in the range of magnetic fields 0-15 mT is the same as obtained by adiabatic correlation to the zero-magnetic-field situation. 
For example, the state correlating to a 
zero-magnetic-field level with $|m_l| = 2$ has contributions $p^{(i)}(m_l)$ mainly from $m_l$ = 2 and -2. In this case of large electric field, the $m_l$ mixing occur 
mainly between the closely spaced $m_l$ and $-m_l$ levels. 

\begin{figure}
 \includegraphics{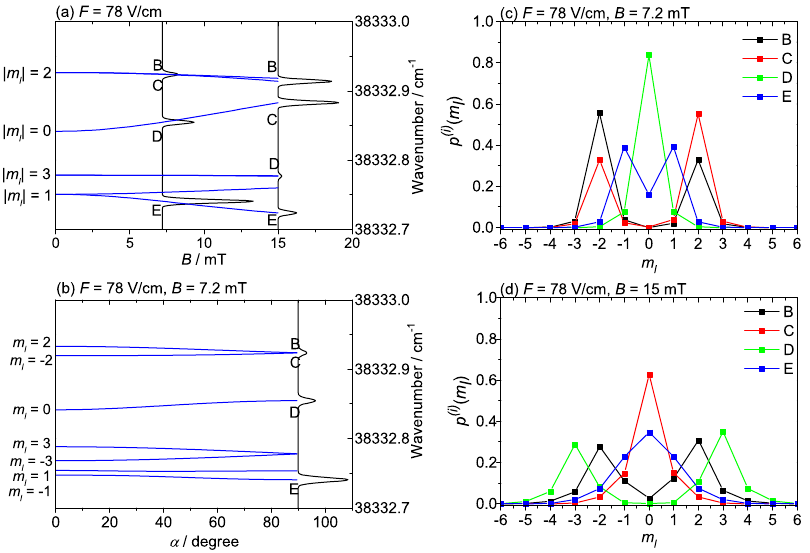}
 \caption{\label{} (a) Enlarged sections of the calculated energy-level structure of $n$ = 30 He Rydberg states for a perpendicular arrangement of magnetic (range 0-15 mT) and electric 
($F$ = 78 V/cm) fields. 
The energy levels are assigned by the magnetic quantum number $|m_l|$ along the electric-field direction in the absence of a magnetic field. (b) Correlation diagram of $n$ = 30 Rydberg states of 
helium for $F$ = 78 V/cm, $B$ = 7.2 mT and  $\alpha$ = 0 - 90$\degree$. The spectrum calculated for $\alpha$ = 90$\degree$ is also displayed. (c-d) Distribution of $m_l$ character corresponding 
to the levels observed experimentally in panel (a) calculated with equation (8).}
 \end{figure}

Because $+m_l$ and $-m_l$ levels are degenerate at zero magnetic field the correlation to $B$ = 0 mT only enables the assignment of the absolute value of $m_l$ using correlation diagrams. 
However, by plotting the level energies against the angle $\alpha$ between electric and magnetic 
fields (as will be discussed in Section 3.5), the transitions can be adiabatically connected to a situation where the electric and magnetic fields are parallel. In this situation, $m_l$ is still a 
good quantum number but $+m_l$ and $-m_l$ are not degenerate. The correlation diagram as a function of $\alpha$ for $F$ = 78 V/cm and $B$ = 7.2 mT is 
shown in figure 12(b) together with the spectrum calculated for $\alpha$ = 90$\degree$. This figure reveals that the final levels of 
the four transitions observed in the spectrum measured at 7.2 mT (see figure 11(e)) can be connected in order of increasing wavenumber to Rydberg states with $m_l$ = -1, 0 and -2 levels.  

\begin{figure}
 \includegraphics{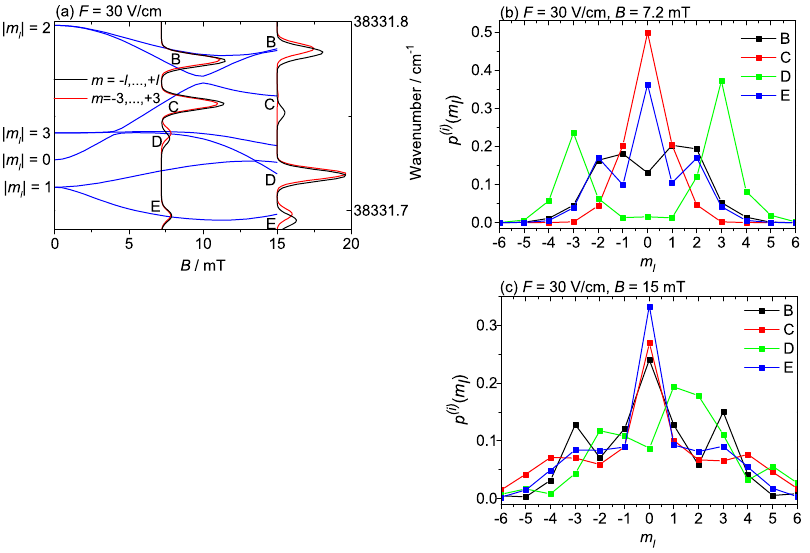}
 \caption{\label{} (a) Enlarged sections of the calculated energy-level structure of $n$ = 30 He Rydberg states for a perpendicular arrangement of magnetic (range 0-15 mT) and electric 
($F$ = 30 V/cm) fields. 
The energy levels are assigned by the magnetic quantum number $|m_l|$ along the electric-field direction in the absence of a magnetic field. Spectra 
shown as red and black lines are spectra calculated at 7.2 and 15 mT using $m_l = -3, ..., +3$ basis and a full $m_l = -l, ..., +l$ basis, respectively. (b-c) Distribution of $m_l$ character 
corresponding to the levels observed experimentally in panel (a) calculated with equation (8).}
 \end{figure}

In the spectra recorded at 30 V/cm (figure 11(a-b)), a repeated structure of four lines can be observed, two of which merge into a single 
line in the high-wavenumber range of both spectra, whereas the weak line in the spectrum measured at 15 mT disappears. The corresponding energy-level diagram is shown in 
figure 13(a) with superimposed calculated spectra for $B$ = 7.2 mT and 15 mT. The spectra calculated including only $m_l = -3, ..., +3$ basis states (shown as red lines) reproduce the spectra 
calculated using the full $m_l = -l, ..., +l$ basis (black lines) reasonably well. The small discrepancies suggest that a mixing with $|m_l|\geq3$ states plays a role at these field 
strengths, especially in the case of the spectrum measured at $B$ = 15 mT and $F$ = 30 V/cm. The distributions of $p^{(i)}(m_l)$ values for the final states observed in the spectra 
recorded at 7.2 and 15 mT and 30 V/cm are presented in figures 13(b) and (c), respectively. 

The spectra calculated at 7.2 and 15 mT and 30 V/cm indicate the importance of transitions to states of $|m_l| = 3$ character and thus provide clear evidence for $m_l$ mixing induced by the 
perpendicular arrangements of the magnetic- and electric-field vectors \cite{korevaar83a}. In this case, the energy separation between levels of different $|m_l|$ values is comparable 
to the splitting between the $m_l$ and $-m_l$ levels, resulting in extensive $m_l$ mixing. This strong $m_l$ mixing prevents one to assign the dominant $m_l$ character 
by adiabatic correlation to the zero-magnetic-field situation. Indeed, the adiabatic correlation 
to the $|m_l|$ zero-magnetic-field level and the largest $|m_l|$ contribution of the $p^{(i)}(m_l)$ distribution are not in agreement. 
For example, the level C in the spectrum recorded at $B$ = 7.2 mT and 30 V/cm adiabatically 
connects to a zero-magnetic-field level with $|m_l| = 2$ but has mainly contributions from $m_l$ = 0 and $\pm1$ (red line in figure 13(b)). 

\subsection{Spectra measured in electric and magnetic fields with arbitrary relative orientations}

\begin{figure}
 \includegraphics{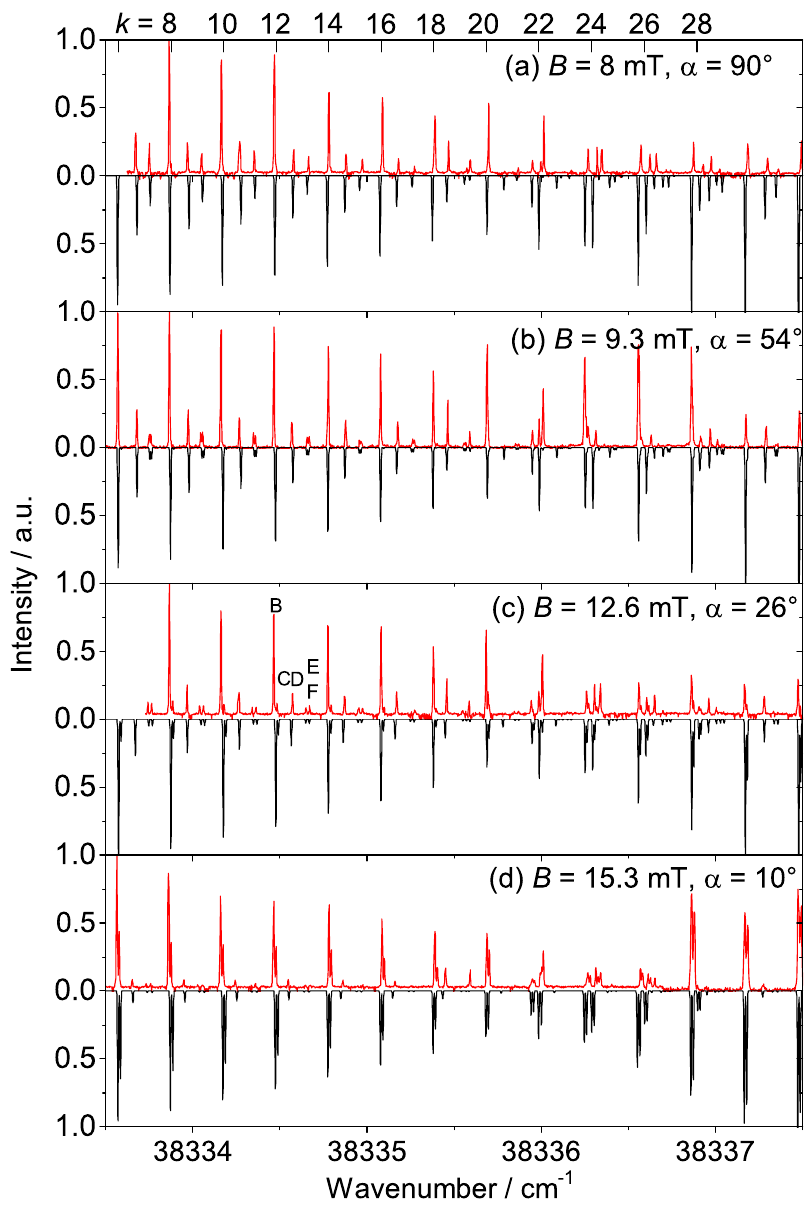} 
 \caption{\label{}Experimental and calculated spectra of $n$ = 30 Rydberg states of helium for several magnetic-field strengths and angles between electric and magnetic fields 
for the electric-field strength $F$ = 78 V/cm. }
 \end{figure}

The $m_l$-mixing processes that take place when the electric- and magnetic-fields vectors are not parallel lead to the observation of more transitions. As illustration, figure 14 shows several spectra of
$n$ = 30 and 31 Rydberg states recorded at an electric field of 78 V/cm for several combinations of magnetic fields and angles. Instead of the regular series of Stark states split in two $m_l = \pm1$
components by the Zeeman effect observed at $\alpha = 0\degree$ (see figure 7(c)), the spectra measured at $\alpha \neq 0\degree$ consist of series of up to five transitions of varying 
strength and spacing, labeled B-F in figure 14(c). The energy-level structure depends on the angle $\alpha$ between the fields. The nonvanishing interaction between states of different nominal $m_l$ 
values leads to more states being optically accessible, to avoided crossings between these states and to $m_l$ changing processes when Rydberg atoms move in regions where the angle 
between the fields varies spatially, as is the case in overlaid electric and magnetic traps (see figure 1). 

In this section, we examine the $\alpha$ dependence of the energy level structure and analyse the resulting $m_l$-mixing and $m_l$-changing processes. To this end, spectra were 
recorded for constant electric- and magnetic-field strengths but variable angles between the field vectors. The angle was adjusted by carefully setting the current 
flowing in the two pairs of coils (see figure 2) and thus the direction of the magnetic-field vector in the $y'$, $z'$ plane while keeping the electric-field vector unchanged. Comparison 
between measured and calculated spectra was used to validate our model, with which specific aspects of the $m_l$-changing processes could then be explored. The calculations 
were made based on Hamiltonian (2). Figure 15 compares the sections of the repeated spectral structures (as for example denoted in figure 14(c) by letters B-F) observed 
for several values of $\alpha$ at field strengths of 78 V/cm and 8 mT (panel (a)) and 30 V/cm and 14.3 mT (panel (b)). A larger 
degree of $m_l$ mixing is expected in the latter case because of the less dominant role of the electric field. This expectation is directly confirmed by the comparison of figures 15(a) and (b). 
The values of $m_l$ used in the following discussion of these spectra always refer to the axis defined by the electric-field vector and is not a good symmetry label when $\alpha \neq 0\degree$.

\begin{figure*}
 \includegraphics{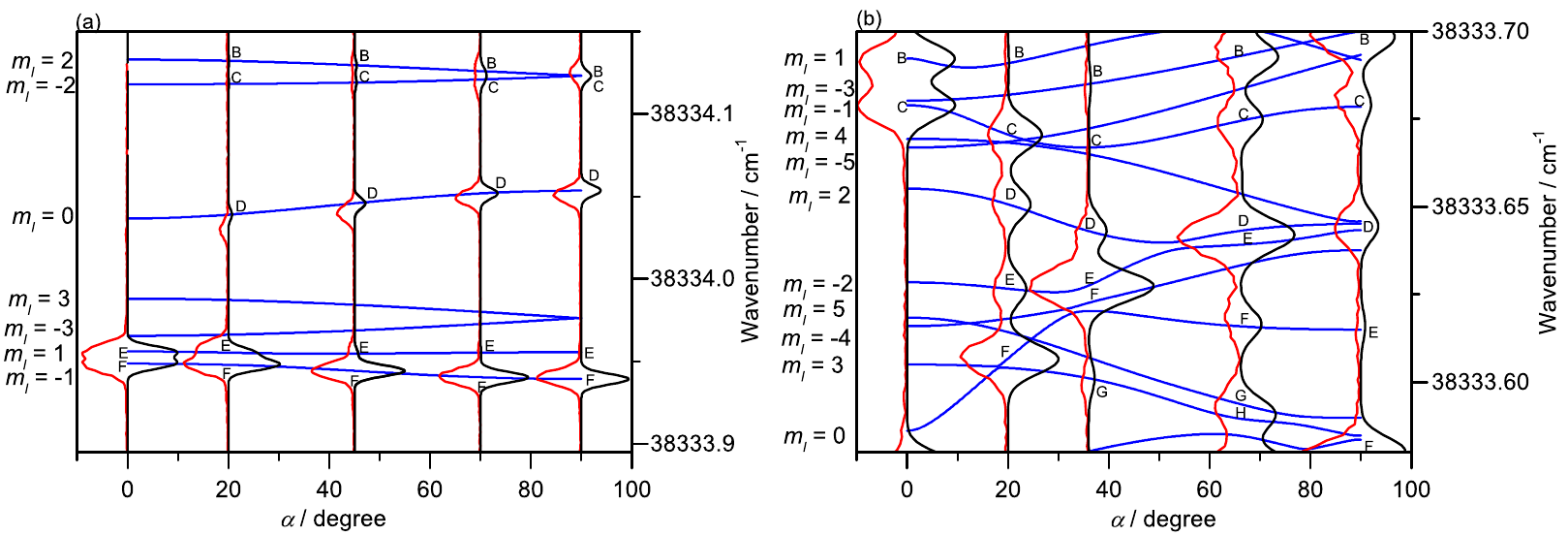} 
 \caption{\label{}Experimental (red lines) and calculated (black lines) spectra of $n$ = 30 Rydberg states of helium for several angles between electric and magnetic fields 
at (a) $F$ = 78 V/cm and $B$ = 8 mT and (b) $F$ = 30 V/cm and $B$ = 14.3 mT. 
The angle dependence of level energies were calculated using Hamiltonian (2) with (a) $|m_l| \leq 3$ and (b) $|m_l| \leq 5$ states included in the basis set. The letters from B to H 
denote a particular final state, which will be referred to in figure 16.}
 \end{figure*}

The $p^{(i)}(m_l)$ values, quantifying the degree of $m_l$ mixing (see equation (8)), corresponding to the spectra recorded at 78 V/cm and 8 mT and presented in figure 15(a) are displayed in 
the left column of figure 16. 
At $\alpha = 0\degree$, $m_l$ mixing does not occur, and the two eigenvectors corresponding to the transitions observed in figure 15(a) have contributions only from $m_l$ = 1 or $m_l$ = -1.
At $\alpha = 20\degree$, all five transitions are to states dominated by a single $m_l$ value corresponding to the zero-magnetic-field situation.
At $\alpha = 45\degree$ and $70\degree$, the final state correlated to $m_l$ = +1 (-1) at zero-magnetic field has significant contribution from $m_l$ = -1 (+1). 
At $\alpha = 90\degree$, only one final level correlating to $|m_l| = 1$ has nonzero intensity in the spectrum with equal contributions from  $m_l$ = +1 and -1. 
The same holds true for final level correlating to $|m_l| = 2$.  Consequently, at 78 V/cm, $m_l$ mixing occurs almost exclusively between the near-degenerate $m_l$ and $-m_l$ levels of the $|m_l|$ pairs. 
The level structures and intensity patterns displayed in figure 15(a) are easiest to interpret near $\alpha = 0\degree$. In this case, each $|m_l| \neq 0$ level 
pair is split in two $\pm m_l$ components separated by $\Delta m_l\mu_BB\cos\alpha$ with  $\Delta m_l =$ 2, 4 and 6 for $|m_l| =$ 1, 2 and 3, respectively. The transitions to $|m_l| =$ 1
levels (lines E and F) are the only ones allowed at $\alpha = 0\degree$. As $\alpha$ increases, transitions to the $m_l =$ 0 (line D) and $|m_l| =$ 2 (lines B and C) levels become observable. 
The separation between the two levels of the $m_l = \pm2$ and $m_l = \pm3$ pairs follows the expected $\cos\alpha$ dependence and decreases with $\alpha$. Surprisingly, the $m_l = \pm1$ 
pair behaves differently: the splitting between the $m_l = \pm1$ components increases with $\alpha$, and only the level correlated with $m_l$ = -1 retains its intensity. With increasing $\alpha$, 
the $m_l = \pm1$ components mix, forming a $c_{-1}|m_l=-1\rangle \pm \sqrt{1-c_{-1}^2}|m_l=+1\rangle$ pair. At large values of $\alpha$, $c_{-1}$ approaches the value of $1/\sqrt{2}$ 
and the negative superposition, which correlates to the $m_l$ = 1 level at $\alpha = 0\degree$, loses its intensity because of the cancellation of transition-dipole amplitudes, 
while its spectral position remains unchanged. The positive superposition is shifted to lower energies and remains strong, but starts sharing its intensity with the $m_l$ = 0 level, 
with which it interacts at $\alpha \neq 0\degree$. The interaction of the negative superposition with the $m_l$ = 0 level vanishes, because of the cancellation. The interactions of the $m_l = \pm1$ 
levels with the $m_l = \pm2$ and $\pm3$ levels are weaker because of the larger spectral separation and because the $|m_l| = 2$ and $3$ levels, unlike the $m_l$ = 0 component, have no "s" character.  

\begin{figure}
 \includegraphics{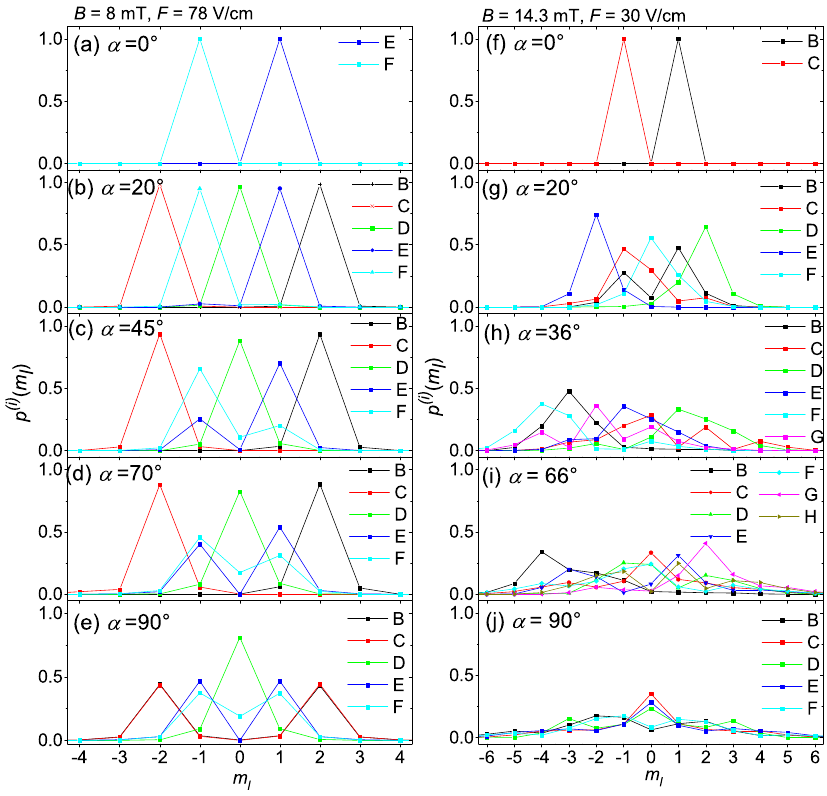} 
 \caption{\label{} Contributions of field-free $m_l$ states corresponding to the spectra presented in figure 15. The left column is for spectra 
measured at $F$ = 78 V/cm and $B$ = 8 mT and the right column for $F$ = 30 V/cm and $B$ = 14.3 mT. The letters from B to H denote a particular final level in figure 15. The points are connected to guide the eye. }
 \end{figure} 

Figure 15(b) shows spectra measured at 30 V/cm and 14.3 mT, i.e., in the regime where the quadratic Stark splittings between the $|m_l|$ levels of the same $k$ value are comparable 
to the pure Zeeman splittings, resulting in a higher degree of $m_l$ mixing. The contributions 
of individual $m_l$ values to the levels observed are shown in the right column of figure 16. 
At $\alpha = 20\degree$, the final states correlating to $m_l$ = +1 (-1) already have contributions from $m_l$ = 0 and 
 $m_l$ = -1 (+1). The final states with dominant $m_l$ = 2, -2 and 0 character (denoted by D, E, F in figures 15(b) and 16) have also contributions from other $m_l$ values. 
The degree of $m_l$ mixing increases with $\alpha$. At $\alpha = 36\degree$, a dominant $m_l$ character can still be determined for each final level and is $m_l$ = -3, -1, 2, 0, -4 and 3 of the 
levels denoted by B, C, D, E, F and G, respectively. Final states in the spectra measured at $\alpha = 66\degree$ and $90\degree$ are so strongly $m_l$ mixed that it is not possible 
to unambiguously identify the dominant $m_l$ character any more. 

\begin{figure}
 \includegraphics{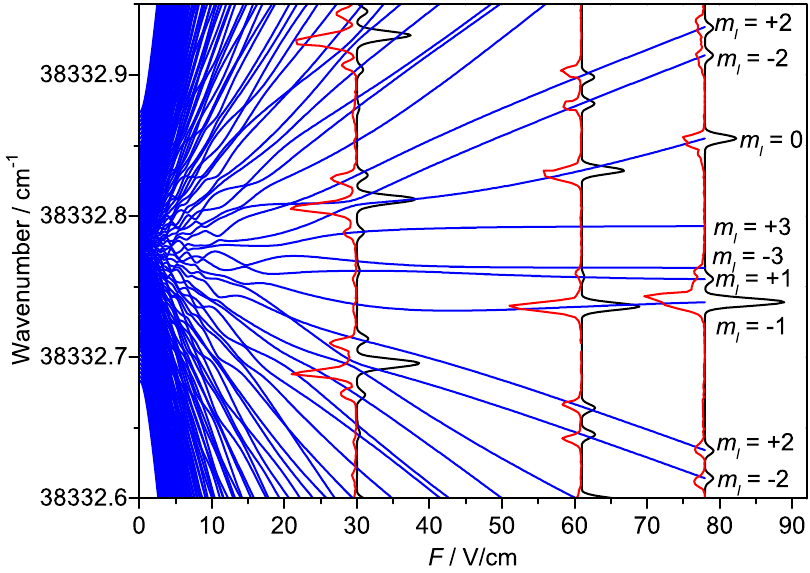}
 \caption{\label{}Experimental (red lines) and calculated (black lines) spectra of ($n$ = 30) helium Rydberg states recorded at $B$ = 13 mT, $\alpha = 34\degree$ and 
(a) $F$ = 30 V/cm, (b) $F$ = 60 V/cm and (c) $F$ = 78 V/cm. Electric-field dependence of level energies of Rydberg states of helium for a fixed $\alpha$ = 34$\degree$ and 
$B$ = 13 mT and for a range of $F$ = 0 - 78 V/cm were calculated using Hamiltonian (2) with (a) $|m_l| \leq 3$ states included in the basis set.}
 \end{figure}

\begin{figure*}
 \includegraphics{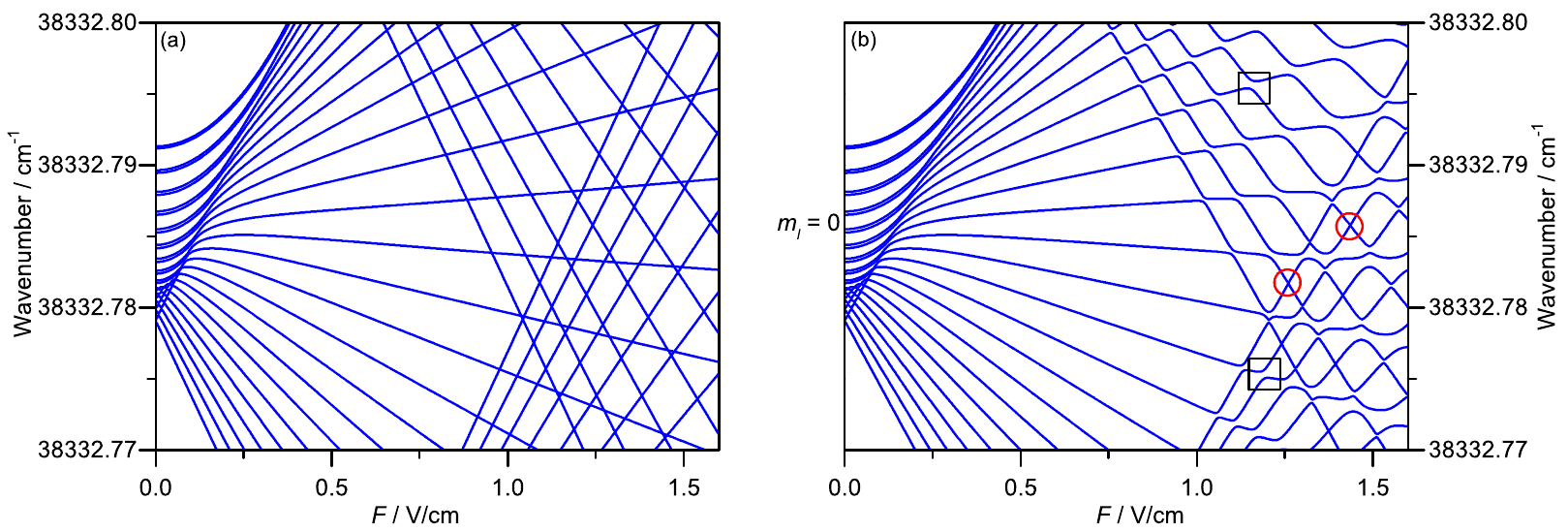} 
 \caption{\label{}Electric-field dependence of $n$ = 30, $S$ = 1 Rydberg states of helium calculated for a magnetic field $B$ = 120 mT and an electric-field range $F$ = 0 - 1.6 V/cm for 
(a) parallel fields ($\alpha = 0\degree$) and (b) $\alpha = 30\degree$. The red circles in panel (b) denote examples of crossings 
between $m_l$ = +1 and -1 levels, whereas the black squares denote examples of avoided crossings between $m_l$ = 0 levels and $m_l$ = +1 or -1 levels .}
 \end{figure*}

The contributions $p^{(i)}(m_l)$ are symmetric around $m_l$ = 0 for all 
three final states in the spectrum measured at $F$ = 78 V/cm, $B$ = 8 mT and $\alpha = 90\degree$. This symmetry is not observed for $\alpha \neq 90\degree$. 
Experimental and calculated spectra of $n$ = 30 Rydberg states of helium at $B$ = 13 mT, $\alpha = 34\degree$ and $F$ = 30, 60 and 78 V/cm are compared in figure 17, which 
also displays the electric-field dependence of the energy levels. The spectra measured at $F$ = 60 and 78 V/cm show the repeated 
structure of the five peaks discussed above (see figure 15(a)). The spectral structure at $F$ = 30 V/cm is dominated by three closely spaced peaks. Figure 17 
illustrates the complex energy-level structure, with multiple avoided crossings also apparent in figures 15(b) and 7(b). Adiabatic traversals of these crossings 
change $m_l$ and often also the electric dipole moment, with consequences for the deceleration and trapping experiments of the kind described in the introduction.

Comparing the energy-level diagrams of $n$ = 30 helium Rydberg states for $\alpha = 0\degree$ and $\alpha = 30\degree$ at $B$ = 120 mT in the range $F$ = 0 - 1.6 V/cm presented in figure 18 nicely 
reveals the effects of $\alpha \neq 0\degree$ on the structure of the $m_l$ = 0 and $m_l$ = +1 and -1 manifolds. The deviation from cylindrical symmetry in the $\alpha$ = 30$\degree$ case 
couples states differing in $m_l$ by $\pm1$, and levels show avoided crossings 
between $m_l$ = 0 levels and $m_l$ = +1 or -1 levels (see the two examples enclosed by black squares in figure 18(b)), which are coupled by this interaction in first order. 
The minimal separation between states differing in $m_l$ by $\pm1$ at the crossings is proportional to the interaction between them and scales as $\sin\alpha$. In contrast, the separation between $m_l$ = +1 and -1 levels at avoided 
crossings (see red circles in figure 18(b)) is very small and not visible on the scale of the figure, because the interaction between them acts only in second order (two steps of $\Delta m_l = \pm1$). 

Consideration of figures 7(b), 16 and 18, however, reveals that most crossings take place at low fields. Consequently, their adverse effect on deceleration and 
trapping could be reduced by lifting the electric-field minimum in the trap to $\sim20$ V/cm, which can be achieved in the quadrupole-like trap used in 
our experiments \cite{hogan08a,seiler11a,hogan11a} by applying a potential difference in the $x$-direction across the end-cap electrodes.  

\section{Conclusions}
The spectroscopy of Rydberg helium (\it n \rm = 30) in a pure magnetic, a pure electric and combined magnetic and electric fields under arbitrary relative
orientations was presented.   
The experimental spectra were recorded in the regime where the Stark and Zeeman interactions are much weaker than the Coulomb interaction but, at the same time, much stronger
than the spin-orbit interaction, which was therefore neglected. The spectra of Rydberg helium in the external fields were 
also calculated by determining the eigenvalues and eigenvectors of the Hamiltonian matrix. All features of the experimental spectra (line positions and intensities) could be 
reproduced well by the calculations, especially in the regions where the levels of the adjacent $n$ manifolds do not overlap. This good agreement enabled us to quantify the 
degree of $m_l$ mixing induced by a nonparallel arrangement of the electric- and magnetic-field vectors and how it is affected by the relative strength of the electric and 
magnetic fields and the angle between them. The degree of $m_l$ mixing is particular large when the spacings between different $|m_l|$ levels of a given $k$ value induced by the quadratic Stark effect is of similar 
magnitude as the spacing between adjacent pure Zeeman levels.  Particular emphasis was placed on the characterization of avoided crossings between Rydberg states in the range of electric 
and magnetic field strengths relevant for magnetic trapping of cold atoms and molecules following Rydberg-Stark deceleration. 
  
\ack
We acknowledge financial support from the Swiss National Science Foundation in the realm of the NCCR QSIT and also under project 200020-149216.


\end{document}